# Presolar Grains in Primitive Ungrouped Carbonaceous Chondrite Northwest Africa 5958

Larry R. Nittler[1], Rhonda M. Stroud[2], Conel M. O'D. Alexander[1], and Kaitlin Howell[1,*]

[1]Department of Terrestrial Magnetism, Carnegie Institution of Washington, Washington, DC 20015, USA. E-mail: lnittler@ciw.edu.

[2]U.S. Naval Research Laboratory Code 6366, 4555 Overlook Ave. SW, Washington DC 20375, USA.

*present address: School of Engineering, École Polytechnique Fédérale de Lausanne, Lausanne, CH-1015, Switzerland

## Abstract

We report a correlated NanoSIMS-transmission electron microscopy study of the ungrouped carbonaceous chondrite Northwest Africa (NWA) 5958. We identified 10 presolar SiC grains, 2 likely presolar graphite grains, and 20 presolar silicate and/or oxide grains in NWA 5958. We suggest a slight modification of the commonly used classification system for presolar oxides and silicates that better reflects the grains' likely stellar origins. The matrix-normalized presolar SiC abundance in NWA 5958 is $18^{+15}_{-10}$ ppm (2σ), similar to that seen in many classes of unmetamorphosed chondrites. In contrast, the matrix-normalized abundance of presolar O-rich phases (silicates and oxides) is $30.9^{+17.8}_{-13.1}$ ppm (2σ), much lower than seen in interplanetary dust particles and the least altered CR, CO and ungrouped C chondrites, but close to that reported for CM chondrites. NanoSIMS mapping also revealed an unusual $^{13}$C-enriched (δ$^{13}$C≈100-200 ‰) carbonaceous rim surrounding a 1.4 μm diameter phyllosilicate grain. TEM analysis of two presolar grains with a likely origin in asymptotic giant branch stars identified one as enstatite and one as Al-Mg spinel with minor Cr. The enstatite grain amorphized rapidly under the electron beam, suggesting partial hydration. TEM data of NWA 5958 matrix confirm that it has experienced aqueous alteration and support the suggestion of Jacquet et al. (2016) that this meteorite has affinities to CM2 chondrites.

**INTRODUCTION**

Presolar stardust grains, with highly anomalous isotopic abundances that indicate origins in the winds or explosions of ancient stars, were discovered as a trace component of chondritic meteorites more than 30 years ago (Zinner et al. 1987). Since then, numerous types of presolar grains – including silicates, oxides, carbides, nitrides and elemental C - have been identified and their laboratory data used to address a wide range of astrophysical and cosmochemical questions (e.g., Zinner 2014; Nittler and Ciesla 2016). Because different types and sizes of presolar grains may respond differently to different processes in the protosolar disk (e.g., evaporation, oxidation, size separation) and in asteroidal parent bodies (e.g., aqueous alteration, thermal metamorphism), it has been long recognized that their abundances in different planetary materials may shed light on these processes (e.g., Huss 1990; Huss and Lewis 1995; Nguyen and Zinner 2004). For example, Huss and Lewis (1995) used noble gas components measured in acid residues of different chondrite classes as tracers of presolar graphite, diamond, and SiC and found abundance variations that could be directly related to thermal processing in the meteorite parent bodies. They also argued that the matrix-normalized presolar grain abundances in the most primitive members of the chondrite groups were roughly CI-like. Presolar grains would not be expected to survive chondrule or refractory inclusion formation, so a correlation between presolar C-rich grain and matrix abundances is not surprising, but their CI-like abundances suggests a genetic relationship between chondrite matrices and the chondrule-free CI chondrites.

The advent of the NanoSIMS ion microprobe at the turn of the century led to the ability to efficiently identify presolar silicates (and other phases) in situ in meteorites and IDPs and in disaggregated meteorites (Messenger et al. 2003; Mostefaoui and Hoppe 2004; Nguyen and Zinner 2004). In addition to the study of the grain compositions and structures to learn about stellar and interstellar processes, much effort has been spent in quantifying the presolar grain inventories in different chondrites, both in an attempt to identify the least altered meteorites and to better understand the effects of the alteration processes themselves (Nguyen et al. 2007; Floss and Stadermann 2009a; Leitner et al. 2012; Floss and Haenecour 2016a; Haenecour et al. 2018).

Broadly speaking, whereas presolar SiC appears to occur with a relatively constant (within a factor of a few) matrix-normalized abundance of a few tens of ppm across all groups of non-metamorphosed chondrites (Davidson et al. 2014), presolar silicate abundances are much more variable, due to their being much more sensitive to both heating and aqueous alteration. A recent compilation of O-rich presolar grain abundances (including both oxides and silicates, though the latter dominate in all but the most altered meteorites) in meteorites and anhydrous interplanetary dust particles (IDPs) collected in the stratosphere can be found in the Appendices of Alexander et al. (2017b). Compared to meteorites, IDPs generally have more presolar silicates and oxides (Messenger et al. 2003; Floss et al. 2006), with an average abundance of 331 ppm, but individual IDPs have abundances ranging up to 1.5 wt.% (Busemann et al. 2009). In contrast, the highest abundance reported to date in a meteorite's matrix is 240±30 ppm for the CO3.0 chondrite Dominion Range (DOM) 08006 (Haenecour et al. 2018; Nittler et al. 2018). We note that Floss and Haenecour (2016b) reported a comparable abundance (275±50 ppm) for the unequilibrated ordinary chondrite (LL) Meteorite Hills 00526, but the data have only been reported in abstract form. Several lightly altered ungrouped and CR carbonaceous chondrites, including Acfer 094, Queen Alexandra Range (QUE) 99177, and Meteorite Hills (MET) 00426, also have relatively high presolar O-rich stardust abundances in the range of 150-200 ppm (Floss and Stadermann 2009b; Nguyen et al. 2010). Other CO, CR, CM, and ordinary chondrites show lower abundances that are commonly argued to reflect destruction during parent-body processing. Moreover, in many chondrites there are clear differences in presolar silicate abundances between interchondrule matrix and fine-grained chondrule rims (Leitner et al. 2016; Haenecour et al. 2018).

Northwest Africa (NWA) 5958 was found as a single 286 g meteorite in Morocco in 2009 and originally classified as an ungrouped carbonaceous chondrite of type 3.0. Preliminary investigations reported in conference abstracts (Ash et al. 2011; Bunch et al. 2011) described NWA 5958 as a 'uniquely primitive' carbonaceous chondrite with bulk CI-like elemental composition, a bulk $^{16}$O enrichment and anhydrous mineralogy. An unaltered CI-like chondrite dominated by matrix would be expected to contain abundances of presolar grains as high as observed in the least altered chondrites studies so far, perhaps approaching the levels seen in IDPs. We thus targeted this meteorite for a dedicated NanoSIMS search, but as described in this paper, found it to have a relatively low abundance of presolar silicates and oxides, as well as petrographic evidence for aqueous alteration (Nittler et al. 2012; Stroud et al. 2014). Subsequently, Jacquet et al. (2016) reported a detailed mineralogical, chemical, and isotopic study of NWA 5958. Their results did not confirm the preliminary reports of Bunch et al. (2011) and Ash et al. (2011), but rather supported that it is a weakly altered ungrouped chondrite bearing some similarities to the CM group, but is most likely from a different parent body. Here we describe in detail our NanoSIMS-based characterization of the presolar grain inventory of NWA 5958. We also report transmission electron microscopy (TEM) data for two presolar grains and their surrounding matrix. Our data are consistent with the conclusion of Jacquet et al. (2016) that this meteorite, while highly primitive, is somewhat altered and possibly related to CMs.

## SAMPLE AND METHODS

A. J. Irving provided a polished thick section of NWA 5958. We performed limited scanning electron microscopy on this section with a JEOL 6500F scanning electron microscope (SEM), equipped with an Oxford Instruments Aztek energy dispersive x-ray spectrometer (EDS) system to identify matrix-rich areas to target for NanoSIMS analysis (Figure 1). Although we did not perform detailed mineralogical or petrographic studies of the NWA 5958 section, it appears similar to that reported for other samples of this meteorite (Jacquet et al. 2016), namely many chondrules a few hundred microns in diameter embedded in a high volume of fine-grained matrix (76% according to Jacquet et al. 2016, similar to that of CM chondrites, Alexander et al. 2007). The matrix areas selected for NanoSIMS analysis were all near the largest chondrule (~2 mm) in the section (Fig. 1) to facilitate easy location in the NanoSIMS.

We used the Cameca NanoSIMS 50L ion microprobe at Carnegie in imaging mode to scan three such areas for their C and O isotopic compositions in three analytical sessions spanning one year (Table 1). A focused ~1 pA $Cs^+$ beam of about 100 nm diameter was scanned over 20 μm × 20 μm regions and 256×256 pixel images of negative secondary ions of $^{12,13}C$, $^{16,17,18}O$, $^{28}Si$, and either $^{30}Si$ (areas A1 and A2) or $^{27}Al^{16}O$ (area A3), together with secondary electrons, were simultaneously acquired. All regions were first pre-sputtered with a much higher-intensity beam to remove the C coat and implant sufficient Cs to achieve stable secondary ion beams. For each region, 20 image frames were acquired with a counting time of 6 ms/pixel, for a total counting time per image of a little over two hours (or ~20 seconds per µm$^2$). After each measurement the stage was automatically moved to an adjacent region for the next analysis. The effective spatial resolution of the images was typically about 120 nm, based on measurements of line profiles across sharp edges.

The images were quantitatively analyzed with the L'image software (L. R. Nittler, Carnegie Institution). Data analysis methods were essentially identical to those described in Nittler et al. (2018). Briefly, images were corrected for the counting system dead time and quasi-simultaneous arrival, and individual image frames aligned and summed. We generated isotopic ratio images ($^{13}$C/$^{12}$C, $^{17}$O/$^{16}$O/, $^{18}$O/$^{16}$O, and $^{30}$Si/$^{28}$Si) and identified presolar grains as regions of interest (ROIs) whose isotopic ratios differed from their surroundings by >4σ, where σ is one standard deviation based on counting statistics. However, most of the reported anomalies are more significant than 5σ. For grains showing isotopic depletions (e.g., in $^{17}$O, $^{18}$O, or $^{13}$C), the counting statistical error was calculated from the total counts expected for a terrestrial isotopic ratio. Grain sizes were determined from the ratio images and corrected for 120-nm beam broadening as described by Nittler et al. (2018). Isotope data for anomalous ROIs were normalized to the average ratios measured in each image. These were assumed to be terrestrial for O, which introduces a negligible error to the results since the O isotopic composition of bulk NWA 5958 is within 10 ‰ of terrestrial (Jacquet et al. 2016). For C, we assumed that the bulk C has a δ$^{13}$C value of -20±20 ‰, comparable to that typically seen in insoluble organic matter in CM and CR chondrites (Alexander et al. 2007).

Three cross-sections, with sizes ranging from 5 µm × 7 µm to 7 µm × 10 µm and centered on selected presolar grains, were extracted from NWA 5958 with in situ focused ion beam (FIB) lift-out with a FEI Nova 600 FIB-SEM at the Naval Research Laboratory (NRL). Analytical TEM studies were performed with a JEOL 2200FS field-emission scanning TEM at NRL, equipped with a Noran System Six EDS spectrometer. Bright-field TEM images were recorded with a Gatan Ultrascan 1000 CCD, with magnification calibrated to a MagICal silicon lattice standard. EDS spectra of individual grains were quantified with Cliff-Lorimer routines, with K factors calibrated from San Carlos olivine and Tanzanian hibonite standards.

## RESULTS

**NanoSIMS measurements**

A total of 38,165 $\mu m^2$, based on $^{16}O$ NanoSIMS images and excluding cracks and large mineral grains, were mapped by NanoSIMS, divided roughly equally among the three areas (Table 1). No anomalies were observed in $^{30}Si/^{28}Si$ ratios, so we limit discussion here to the results for C and O. Example images of area A2-1-29, containing two presolar SiC grains and one presolar O-rich grain are shown in Figure 2.

*Carbon Isotopes*

The NanoSIMS C isotopic mapping identified 12 sub-micron regions with $^{13}C/^{12}C$ ratios significantly different from those of typical carbonaceous regions in the images; 11 are enriched in $^{13}C$ and one is $^{13}C$-depleted (Table 2). Note that the pre-sputter current was set too low for the analytical session on area A1, such that the C coat was not fully removed prior to each measurement. This means that C isotopic data for this run include some contribution from the C coat and $^{12}C/^{13}C$ ratios for the three anomalous grains identified in this area were thus calculated only from the mostly uncontaminated last few frames of the images. Ten of the $^{13}C$-rich grains are identified as presolar SiC, based on correlated signal in the $^{28}Si$ images (e. g., Figure 2). The $^{13}C$-depleted grain and one $^{13}C$-rich grain appear to have very low Si and thus likely are presolar graphite grains. The SiC and presumed graphite grains range in diameter from 150 nm to 540 nm.

We calculated the abundance of presolar SiC in the matrix by dividing the total area covered by the grains (corrected for beam-broadening) by the total analyzed area, excluding cracks. The errors in the abundances were calculated from Poisson statistics, based on the number of grains and the confidence limit tables of Gehrels (1986). The average SiC abundance and abundance ranges corresponding to one- and two-$\sigma$ errors for each of the individual areas and the total data set are given in Table 1 and shown in Fig. 3b. The average SiC abundance for the total data set is $18^{+15}_{-10}$ ppm ($2\sigma$). While there are large variations between the three analyzed areas, these reflect the very low counting statistics and all agree with the average abundance within $2\sigma$ uncertainties. Note that the insufficient pre-sputtering of area A1 could possibly decrease the detection efficiency for C-anomalous grains in this area, but this is not evident in the abundance data.

In addition to the C-anomalous presolar grains, the NanoSIMS C mapping revealed an unusual 300-nm thick carbonaceous rim with a clear enrichment in $^{13}$C (average $\delta^{13}$C=125±10 ‰, with some pixels reaching 250 ‰, Figure 4) surrounding a 1.4 µm diameter silicate grain. If this was a uniform coating around a spherical grain, the total amount of carbonaceous material in the rim would be about 3.4 µm$^3$ or the equivalent of a 2-µm diameter carbonaceous grain. An SEM-EDS measurement of the rimmed silicate grain found it to be compositionally homogeneous for O, Mg, Al, Si, and Fe. Standardless EDS quantification gives a composition of Mg$_{1.2}$Al$_{0.4}$Fe$_7$Si$_2$O$_{13}$, most likely a phyllosilicate (e.g., cronstedtite or serpentine+magnetite). We attempted to prepare a FIB section of this grain and its rim for analysis by TEM, but the section was unfortunately lost during preparation.

***Oxygen Isotopes***

We identified a total of 20 presolar O-rich grains in the NanoSIMS images (Table 3). As with all NanoSIMS in situ studies (see, e.g., Nguyen et al. 2010), the derived O isotopic ratios are likely affected to varying degrees by contributions of signal from surrounding material due primarily to tails on primary ion beams but possibly also to re-deposition of sputtered atoms. This isotope dilution shifts measured isotopic compositions towards normal and is the most likely explanation for the narrower distribution of $^{18}O/^{16}O$ ratios observed in presolar silicates compared to presolar oxides (Nittler et al. 2018), most of which have been measured as separated grains, not in situ. The O-anomalous grains range in diameter from 180 nm to 400 nm, with an average diameter of 277 nm. All but one had $^{28}Si^-/^{16}O^-$ secondary ion ratios within ≈50% of the bulk average ratio observed in the same images, suggesting most are silicates. The exception, $^{17}O$-depleted grain A3-2-14 has a $^{28}Si^-/^{16}O^-$ ratio more than an order of magnitude lower than the surrounding matrix and is likely an oxide. However, the O isotopic anomaly was not correlated with signal in the $^{27}Al^{16}O^-$ image either, so it is unlikely to have been an Al-oxide. Unfortunately, this grain could not be re-located by SEM. In any case, as long recognized (Nguyen et al. 2010), NanoSIMS secondary ion ratios are not always reliable for phase determination of presolar O-rich grains found in situ, even for discriminating silicates from oxides. For example, as discussed further below, one of the grains with a matrix-like $^{28}Si^-/^{16}O^-$ ratio, A2-2-15, was subsequently determined by TEM analysis to be Al-Mg spinel.

The O-isotopic data for the twenty presolar O-rich grains found in NWA 5958 are compared to literature data in Figure 5; the data fall within the previously observed ranges of O isotopes in presolar oxides and silicates. Fifteen of the new presolar grains are enriched in $^{17}O$ and have terrestrial to slightly sub-terrestrial $^{18}O/^{16}O$ ratios and thus belong to the dominant Group 1 population (Nittler et al. 1997) of presolar oxides and silicates. Note, however, that it is possible that some of these grains, especially those with $^{17}O/^{16}O$ ratios $<1\times10^{-3}$ may in fact be more $^{18}O$-depleted Group 2 grains affected by isotopic dilution from surrounding material. Of the remaining five, three are $^{18}O$-enriched Group 4 grains, two of which are $^{17}O$-depleted, and two are Group 3 grains with $^{17}O$-depletions and roughly terrestrial $^{18}O/^{16}O$ ratios.

The abundance of presolar O-anomalous grains in NWA 5958 was determined by the same procedure described above for SiC. We note that Nittler et al. (2018) reported a Monte Carlo method to more accurately determine errors on presolar grain abundances that takes into account grain sizes and this was used here to calculate the errors on the total O-rich presolar grain abundance. However, we found that the algorithm gave essentially identical results to those obtained via the number of grains and the Gehrels (1986) tables for the individual area abundance estimates, so we used the latter here for individual areas. The average abundance and abundance ranges corresponding to one- and two-σ errors for each of the individual analyzed areas and the total data set are given in Table 1 and shown in Figure 3a. The average abundance is $30.9^{+17.8}_{-13.1}$ ppm (2σ) and, as for SiC, the abundances measured in the three individual areas agree with the average within uncertainty (Fig. 3a).

**Transmission electron microscopy**

Three matrix cross-sections containing three O-rich presolar grains were extracted from NWA 5958, but only two of the presolar grains extended far enough below the surface to preserve a useful volume for mineralogical study. TEM analysis of the matrix (Fig. 6) reveals mineralogy characteristic of CM chondrites (Brearley and Jones 1998), dominated by intergrown serpentines with fern and platy textures, along with tochilinite-rich Fe, Ni, S, O assemblages (also referred to as type-I PCP), some nominally anhydrous silicates, and isolated Fe oxides and Fe,Ni sulfides. A chromite and Fe,Ni phosphide assemblage similar to those previously reported in CMs (Bunch et al. 1979; Devouard and Buseck 1997), was also observed close to the tochilinite.

Presolar O-rich grain A2-1-5 (previously named NWA e1-5; Stroud et al. 2014) was identified as enstatite (Fig. 7). EDS measurements indicated it was very near stoichiometric $MgSiO_3$: $Mg_{0.961}Al_{0.013}Ca_{0.003}Cr_{0.005}Mn_{0.002}Fe_{0.004}Si_{1.001}O_3$. The grain exhibited lamellar contrast indicative of clino-orthopyroxene intergrowth in bright-field scanning TEM (STEM) imaging (Fig. 7A) and a Fast Fourier Transform diffractogram of a high-resolution TEM image (Fig. 7C) shows reflections with d-spacings corresponding to clinoenstatite (1 5.05 Å moire fringe; 2 3.00Å (-221); 3 2.32 Å (002), (221), (-231); 4 2.06 Å (-331)). However, the latter reflections degraded during further high-resolution imaging, indicating rapid amorphization under the electron beam. This rapid susceptibility to amorphization is most likely due to partial incorporation of water during parent body hydration.

Grain A2-2-15 (previously named NWA e2-15; Stroud et al. 2014) was identified as Mg-Al spinel with minor Cr ($Mg_{0.97}Cr_{0.11}Al_{1.91}O_4$), on the basis of STEM-EDS, electron diffraction, and high-resolution TEM lattice images (Fig. 8). The surrounding matrix shows the characteristic fern and platy serpentines textures, as well as nanoscale Fe, Ni sulfides and oxides characteristic of CM2s.

## DISCUSSION

**Petrography of NWA 5958 matrix**

The TEM analysis of three FIB sections extracted from the matrix of NWA 5958 provide strong confirmation at the sub-micrometer scale of the previous report based on electron microprobe analysis that NWA 5958 has affinities to the CM group, and is better classified as petrographic grade 2 than 3 (Jacquet et al. 2016). Specifically, the TEM data reveal that the matrix is dominated by platy and fern textured serpentines while silicate glass is absent. Some nominal anhydrous silicates are present, as are PCP assemblages with intergrown sulfide and oxide components. The presence of an unusual chromite- phosphide assemblage, previously reported only for CMs, is also particularly diagnostic.

**Origins of Presolar Grains**

The extremely unusual isotopic ratios of presolar grains indicate that they formed in the winds and explosions of evolved stars, but identifying the type of stellar source for a given grain depends on comparisons of astronomical data and theoretical predictions with multiple isotopic ratios measured in the grains. For example, presolar SiC grains have been classified into several groups based on their C, N, and Si isotopic compositions (Zinner 2014), with >90% belonging to the dominant "mainstream" population. Mainstream grains have $^{12}C/^{13}C$ ratios in the range of 15-100, and N, Al-Mg, and trace-element isotopic compositions indicating an origin in low-mass (1.5-3 $M_\odot$) C-rich asymptotic giant branch (AGB) stars. Type A+B grains, making up 4-5% of all SiC, have $^{12}C/^{13}C$ ratios lower than 10, as do the much rarer highly $^{15}N$-enriched C2 and nova grains (Liu et al. 2016). A+B grains likely had multiple origins, including J-type carbon stars and supernovae (Liu et al. 2017a; Liu et al. 2017b). The remaining few % of SiC grains belong to various other groups with origins in either supernovae (C1, X) or low-metallicity AGB stars (Y, Z). For the presolar SiC grains identified here in NWA 5958, we have only $^{12}C/^{13}C$ ratios and it is thus not possible to uniquely assign stellar sources. Nevertheless, on statistical grounds alone, it is most likely that the ten grains with $^{12}C/^{13}C$ between 20 and 69 (Table 2) are mostly mainstream grains from AGB stars and the two grains with $^{12}C/^{13}C$ <10 are probably AB grains, though further pinning down their origins is not possible with the limited data.

Similarly to presolar SiC, presolar oxides were classified into groups, on the basis of their O isotopic ratios (Fig 5a, Nittler et al. 1997). Since presolar silicates were discovered, it has also become clear that they strongly overlap in their O isotopic compositions with the presolar oxides and are generally also divided into the same groups. A large majority of both presolar oxides and silicates, >75%, belong to Group 1 and have $^{17}O$ enrichments and $^{18}O/^{16}O$ ratios ranging from about half-solar to solar. These compositions match astronomical observations of and theoretical predictions for the envelopes of evolved O-rich AGB stars (Smith and Lambert 1990; Boothroyd et al. 1994). It has been shown that the Group 1 grains largely originated from such stars, with masses up to ≈2.2 $M_\odot$ (Nittler et al. 1997, 2008; Nittler 2009) and this is the most probable origin for the 15 Group 1 grains identified in this study as well.

However, when considering other Groups, there are notable differences between the isotopic distributions of the two mineralogical classes of grains. A much larger fraction of presolar oxides (10%) belong to Group 2, with $^{18}O/^{16}O$ ratios smaller than $1\times10^{-3}$, than of the silicates (2.5%). This almost certainly reflects the effect of isotopic dilution on the in situ measurements of presolar silicates, as this will more easily mask large isotopic depletions than enrichments (e.g., Nguyen et al. 2010). Group 3 grains were defined by Nittler et al. (1994) as those with depletions in both $^{17}O$ and $^{18}O$, relative to terrestrial O isotopes. This may also explain the lower abundance of Group 3 grains among the silicates (3%) compared to the oxides (5.5%). However, in addition to being slightly less numerous than Group 3 oxides, Group 3 silicates show a distinct composition; whereas most of the Group 3 oxides have large $^{18}O$-depletions, the Group 3 silicates are shifted to higher $^{18}O/^{16}O$ ratios. Similarly, whereas most Group 4 oxides have both $^{17}O$- and $^{18}O$ enrichments, Group 4 silicates tend to have lower $^{17}O/^{16}O$ ratios, with many falling in the $^{18}O$-enriched, $^{17}O$-depleted quadrant of the O 3-isotope plot (Fig. 5), where few oxides plot. These differences are reflected in the $^{17}O/^{18}O$ ratios of the grains: the median value of this ratio for Group 3 and 4 oxides is 0.19, 27% higher than that of the Group 3 and 4 silicates (0.15). The five Group 3 and 4 presolar grains found in NWA 5958 all fall in the $^{17}O$-depleted range of presolar silicates, compared to presolar oxides.

It has been long argued that most of the Group 3 oxide grains originated in low-mass and low-metallicity AGB stars (Nittler et al. 1997, 2009), with their lower-than-solar $^{17}O/^{16}O$ and $^{18}O/^{16}O$ ratios reflecting the chemical evolution of the Galaxy (GCE) since the low-mass parent stars of the grains would have formed early in the Galaxy (Nittler and Cowsik 1997), and these ratios would have then been expected to increase with time. However, almost all of the Group 3 silicates, including the $^{17}O$-depleted grains identified here, plot to the $^{18}O$-rich side of the predicted line for O isotopes GCE (Fig. 5b), and it is not expected that low-metallicity AGB star evolution would lead to such envelope compositions. Rather, it has been suggested (Nittler et al. 2008) that presolar grains with these compositions are related to the Group 4 grains, and have an origin in Type II supernovae (SN). Although most of the O ejected in SN is almost-pure $^{16}O$, only two presolar oxide grains have been found with such a composition (Nittler et al. 1998; Gyngard et al. 2010). In contrast, SN oxides and silicates are largely $^{18}O$-rich (Group 4) and their O is dominated by material from the H-rich stellar envelope of the pre-SN massive star and the He-burning, C-rich He/C shell, with small contributions from interior zones (Choi et al. 1998; Messenger et al. 2005; Nittler et al. 2008). This is supported by multi-element (e.g., Mg, K, Ca, Fe) isotopic data for several Group 4 oxide and silicate grains (Zinner et al. 2005; Nittler et al. 2008; Nguyen and Messenger 2014).

We show as long-dashed lines on Fig. 5b mixing lines, calculated from a 15 M$_\odot$ SN model of (Woosley and Heger 2007), for mixing between the He/C zone and the H-rich envelope, and mixing between the He/C zone and $^{16}$O-rich inner zones, and as dashed-dot lines mixing between the $^{16}$O-rich zones and two different mixtures of envelope and He/C zone material. Clearly, almost all Group 4 and Group 3 grains plotting to the right of the GCE line, including the five Group 3 or 4 NWA 5958 grains, may be explained by such mixtures, and SNe are thus plausible sources. Isotopic data for additional elements besides O for more $^{18}$O-enriched Group 3 and $^{17}$O-depleted Group 4 grains are sorely needed, however, to test this hypothesis. If all of these grains are indeed from SNe, then SN grains make up about 14% of presolar silicates and 9% of presolar oxides.

We thus propose a slight modification of the Group definitions of presolar silicates and oxides as defined by Nittler et al. (1994, 1997), indicated on Fig. 5a, to better reflect the likely stellar sources of the Group 3 and 4 grains: We suggest the Group 3 label be reserved for grains with depletions in both $^{17}$O and $^{18}$O and with $^{18}$O/$^{17}$O less than the solar value of 5.2. In this scheme, all grains with $^{18}$O/$^{17}$O larger than the solar value would be called Group 4.

**Presolar Grain Abundances**

NWA 5958 matrix contains, on average, $18^{+15}_{-10}$ ppm (2$\sigma$) presolar SiC and $30.9^{+17.8}_{-13.1}$ ppm (2$\sigma$) presolar silicates and oxides (Table 1, Figure 3). The SiC abundance is in the range seen for most unmetamorphosed C chondrite groups (Huss and Lewis 1995; Davidson et al. 2014), but the abundance of presolar O-rich grains is considerably lower than that seen in both IDPs and many of the least altered C chondrites (Alexander et al. 2017b). We compare the derived presolar grain abundances in NWA 5958 to those of other chondrites in Fig. 9. The least altered CR chondrites (MET 00426, QUE 99177), the ungrouped Acfer 094, and the CO3.0 meteorites DOM 08006, ALHA 77307 and LAP 03117, all have similar SiC abundances of 10-50 ppm, and high silicate and oxide abundances of 175-240 ppm. Two CR2 chondrites, NWA 852 and GRV 021710 have had substantially higher SiC abundances reported on the basis of in situ NanoSIMS searches (Leitner et al. 2012; Zhao et al. 2013). The derived SiC abundances for these two meteorites are based on relatively small examined areas and are in fact within 2-$\sigma$ errors of the range of other CRs. Moreover, in the case of GRV 021710 the high abundance is largely driven by a single large grain. Thus, these unusual high abundances may be statistical flukes, but additional studies are warranted.

There are few studies reporting on presolar silicate and oxide abundances in CM chondrites, and only in abstract form (Mostefaoui 2011; Heck et al. 2013; Leitner et al. 2019; Verdier-Paoletti et al. 2019); all have found significantly lower abundances than seen in the least-altered CRs, COs, and Acfer 094. The abundance in NWA 5958 is slightly higher than but within uncertainty of the value determined for fine-grained chondrule rims in six CM chondrites (18±5 ppm, 1$\sigma$, Figure 9; Leitner et al. 2019) and much higher than that reported for Paris (~10 ppm, Mostefaoui 2011; Verdier-Paoletti et al. 2019), reported to be the least-altered CM (Hewins et al. 2014). The NWA 5958 SiC abundance is very similar to the average value of 22±5 ppm found for CMs by Davidson et al. (2014).

The similarity in presolar grain abundances, especially of silicates and oxides, in NWA 5958 to those reported for CM chondrites provides further support for this meteorite being related to CMs, already suggested based on petrography and bulk chemistry (Jacquet et al. 2016). It is clearly evident that presolar silicates and oxides are in lower abundance in CM chondrites, even relatively less-altered ones, compared to CRs and COs. This may indicate that there was a heterogeneous distribution of presolar O-rich grains in the C chondrite forming region of the protosolar disk, perhaps due to variable nebular destruction. However, recent studies have provided evidence that most CMs are more hydrated than most CRs, based on higher water contents (determined from bulk H and C contents; Alexander et al. 2013) and higher phyllosilicate abundances (determined from x-ray diffraction; Howard et al. 2015). We thus consider the lower abundance of O-rich presolar grains in NWA 50958 and CM chondrites most likely to be further evidence of this, rather than a reflection of nebular processes.

**Presolar Grain Microstructures**

The level of hydrothermal alteration present in the matrix is severe enough that it has likely affected the preservation of O-rich presolar grains, leading to preferential loss of amorphous silicates. Grain A2-1-5 (Fig. 7), an enstatite grain that amorphized rapidly under the electron beam, appears to have been moderately hydrated to the point of increased susceptibility to radiation damage, but not so severely that O isotope exchange prevented its identification as a presolar grain. This suggests that the measured O isotope composition is intermediate between its original circumstellar composition and solar values. For example, A2-1-5, with an $^{17}O/^{16}O \approx 8 \times 10^{-4}$, could have formed with as much as six times more $^{17}O$ and still be in the Group 1 field (Fig. 5a).

Astronomical observations of AGB stars suggest that enstatite dominates over forsteritic olivine among the crystalline silicates they produce, but olivine-like compositions dominate amorphous silicates (e.g., Demyk et al. 2000; Molster et al. 2002). A compilation of Auger spectroscopic measurements of hundreds of presolar silicates indicates that about 20% have "pyroxene-like" and a similar fraction has "olivine-like" Mg(+Fe, Ca)/Si ratios (Floss and Haenecour 2016a; Floss 2018), with the remainder having ratios inconsistent with either class of silicates. Of the >50 presolar silicate grains analyzed to date by TEM techniques (see Floss and Haenecour 2016a for a recent compilation of literature results), roughly one third is crystalline and these are dominated (~2/3) by olivine. Only a handful of presolar enstatite grains, including A2-1-5, and one $MgSiO_3$ grain with perovskite structure have been reported (Vollmer et al. 2007, 2013; Nguyen et al. 2016). Thus, the crystalline presolar grain data (which are predominately for grains from AGB stars) are in conflict with astronomical data indicating that more enstatite than forsterite is produced. The typical grain size of presolar silicates analyzed by TEM of a few hundred nm overlaps with the size distribution believed to exist for circumstellar dust around AGB stars, so the discrepancy is unlikely to be due to sampling different size distributions. It is also unlikely that enstatite is substantially more susceptible to destruction by interstellar or solar system processes than is olivine. Since the presolar grain data are based on direct laboratory measurements of compositions and microstructures of individual grains, and the astronomical results are based on spectral fitting methods which rely on numerous parameters and assumptions about size distributions, grain shapes, etc., we suggest that the source of the discrepancy most likely lies in the interpretation of astronomical spectra from O-rich AGB stars.

The presolar Al-Mg spinel with minor Cr (A2-2-15, Fig. 8) was apparently much less affected by aqueous alteration than the enstatite grain A2-1-5, with no sign of increased sensitivity to the electron beam analysis. Comparison of Al and Fe EDS maps shows that it is enclosed in Fe-rich matrix material, and it is possible that the minor Cr in the spinel reflects hydrothermal alteration, rather than the original condensation composition. The presence of Cr in presolar spinel is not unusual, however (Zinner et al. 2005; Zega et al. 2014). Limited TEM data have been reported previously for presolar spinel grains (Zega et al. 2014) and grain A2-2-15 is not markedly different from other Al-Mg spinels in size, chemistry, or structure, though one composite spinel-silicate grain has been reported (Nguyen et al. 2014) and the structure of a larger presolar Fe-Cr spinel grain was distinctly polycrystalline (Zega et al. 2014).

**C-anomalous Grain Rim**

The origin of the unique carbonaceous rim around a silicate grain shown in Fig. 4 is mysterious. The isotopic composition of the carbonaceous material ($\delta^{13}C$=100-200 ‰) is in the range of rare carbonaceous grains observed before in NanoSIMS surveys of the most primitive chondrites and often associated with either enrichments or depletions in $^{15}N$, relative to terrestrial $^{15}N/^{14}N$ ratios (Floss and Stadermann 2009b; Bose et al. 2012; Nittler et al. 2018). Theoretical astrochemical models predict large fractionations in light elements like H, C, and N in molecular clouds (Ehrenfreund and Charnley 2000) and thus an interstellar origin has been invoked for the $^{13}C$-rich organic materials found before in meteorites (Floss and Stadermann 2009b; Bose et al. 2012, 2014). However, the previously reported anomalous materials appeared as isolated sub-µm grains, not as an apparent rim or coating around a mineral grain as seen here. Moreover, based on the SEM-EDS compositional measurement, the rimmed grain is probably a phyllosilicate, almost certainly formed by aqueous alteration in the meteorite's parent asteroid.

The rimmed grain may have formed as a circumstellar or interstellar anhydrous silicate grain coated in porous $^{13}C$-rich organic matter. During aqueous alteration of the parent body, fluids altered the grain to a phyllosilicate without significantly affecting the C isotopic composition of the carbonaceous rim. Note that the enclosed silicate has O isotopic ratios identical to the surrounding matrix within ~1%. Thus, if this explanation is correct, either the presolar grain had an O isotopic composition that was extremely close to that of bulk C chondrites or the alteration completely equilibrated its isotopic composition. However, given that this object is vastly larger than typical interstellar grains (and larger than any reported presolar silicate), we think this explanation is unlikely.

We suggest instead that the carbonaceous material formed separately from the silicate prior to the two materials coming together. The carbonaceous rim corresponds to a grain of about 2 μm in diameter. This is larger than typical carbonaceous grains in C chondrite matrices, but larger inclusions and veins are also seen in C chondrite matrices (Peeters et al. 2012; Le Guillou et al. 2014; Alexander et al. 2017a), some with highly anomalous H and/or N isotopic compositions, and large aggregates of highly anomalous material are also seen in ultracarbonaceous Antarctic micrometeorites (Duprat et al. 2010). One possibility is that during compaction of the parent body, a phyllosilicate grain (or its precursor) was pushed into a preexisting interstellar carbonaceous grain and the meteorite section was serendipitously cut in such a way that the resulting object appears as a rimmed grain. While possible, this would require a remarkable coincidence to result in an object with a rim of such uniform apparent thickness. Alternatively, the C coating may have been deposited on the silicate from a fluid during aqueous alteration. It is unfortunate that the rimmed grain was lost, as additional data (H and N isotopes, microstructure, chemical functionality) may have provided crucial clues that would have allowed us to infer the origin of this unusual object.

## Conclusions

We have conducted a NanoSIMS isotopic survey and follow-up TEM investigation on a polished section of the primitive ungrouped C chondrite NWA 5958. From our data, we can draw the following conclusions:

1. We identified 10 presolar SiC grains, 2 likely presolar graphite grains, and 20 presolar silicate and oxide grains in NWA 5958, all with isotopic ratios in the ranges of previously identified presolar grains. The SiC grains and 15 of the O-rich grains most likely originated in low-mass asymptotic giant branch (AGB) stars and 5 O-rich grains probably condensed in the ejecta of Type II supernovae. We have also proposed a slight modification of the presolar O-rich grain classification system of Nittler et al. (1994 1997): grains with $^{18}O/^{17}O$ ratios higher than solar, even those with depleted $^{17}O/^{16}O$ ratios, are likely from supernovae and thus classified as Group 4 in the new system.

2. The matrix-normalized presolar SiC abundance in NWA 5958 is $18^{+15}_{-10}$ ppm (2σ), based on 10 identified grains. This is in the range seen for presolar SiC in many classes of unmetamorphosed meteorites. In contrast, the abundance of presolar O-rich phases (silicates and oxides) is $30.9^{+17.8}_{-13.1}$ ppm (2σ), much lower than seen in IDPs and the least altered CR, CO and ungrouped C chondrites, but close to that reported for CM chondrites.

3. We obtained TEM data on two presolar O-rich grains with isotopic compositions indicating AGB star origins. One was found to be crystalline enstatite that rapidly amorphized under the electron beam, likely due to limited hydration on the meteorite parent body. The other was Al-Mg spinel with minor Cr, and appeared similar to spinels previously studied by TEM.

4. TEM observations of three FIB liftout sections of NWA 5958 matrix indicated the presence of phyllosilicates and other phases commonly seen in CM2 chondrites, providing strong confirmation at the sub-micrometer scale to the conclusions of Jacquet et al. (2016), who found that NWA 5958 has affinities to the CM group, and is better classified as petrographic grade 2 than 3.

5. We identified an unusual $^{13}C$-enriched ($δ^{13}C≈100$-$200$ ‰) carbonaceous rim surrounding a 1.4 μm diameter phyllosilicate grain. While the isotopic anomaly probably reflects a heritage in presolar, interstellar molecular cloud chemistry, exactly how and when the rimmed grain formed is unclear.

In contrast to initial reports, and as was concluded in the detailed petrographic and chemical study of Jacquet et al. (2016), NWA 5958 is not a "uniquely primitive" meteorite with CI-like bulk chemistry and no parent-body alteration. The presence of phyllosilicates and the relatively low abundance of presolar silicates and oxides clearly indicates that this meteorite has seen some aqueous alteration. Moreover, the similar abundance of presolar O-rich grains in this meteorite to those reported for CM chondrites provides further support to NWA 5958 being related to this meteorite group.

*Acknowledgements*: This work is dedicated to the memory of Dr. Christine Floss, a friend, colleague, and leader in the study of presolar silicate grains in meteorites. We thank Tony Irving for providing the sample of NWA 5958 for this study and Drs. Pierre Haenecour and Ann Nguyen for helpful reviews. This work was supported by NASA through grants NNX10AI63G and NNX17AE28G to LRN and NNH13AV45I and 80HQTR190038 to RMS.

Table 1. Analyzed areas and presolar grain abundance results for NWA 5958.

| Area | Analysis Date | Area Analyzed ($\mu m^2$) | Presolar SiC Grains | | | | Presolar O-rich Grains | | | |
|---|---|---|---|---|---|---|---|---|---|---|
| | | | # Grains | Abundance (ppm) | 1σ range | 2σ range | # Grains | Abundance (ppm) | 1σ range | 2σ range |
| A1 | 7-2011 | 11,950 | 3 | 8.5 | 3.8 - 16.9 | 1.6 - 25.1 | 2 | 14.5 | 5.1 - 33.5 | 1.7 - 53.1 |
| A2 | 4-2012 | 13,130 | 6 | 41 | 24.9 - 65.9 | 14.8 - 90.9 | 11 | 56 | 39.4 - 78.6 | 27.6 - 101.3 |
| A3 | 7-2012 | 13,090 | 1 | 2 | 0.3 - 6.7 | 0.05 - 11.5 | 7 | 21 | 13.1 - 31.9 | 8.2 - 43.2 |
| Total | | 38,170 | 10 | 18 | 12 - 25 | 8 - 33 | 20 | 30.9 | 24.5 – 40.0 | 17.8 - 48.7 |

Table 2. C-anomalous presolar grains in NWA 5958

| Grain[a] | Diameter[b] (nm) | $\delta^{13}C/^{12}C$ (‰, ±1σ) | $^{12}C/^{13}C$ (±1σ) |
|---|---|---|---|
| SiC-A1-1-3 | 145 | 1,700±200 | 33±2 |
| SiC-A1-2-1 | 210 | 8,920±330 | 9.0±0.3 |
| SiC-A1-2-28 | 265 | 1,570±220 | 35±3 |
| SiC-A2-1-12 | 300 | 377±65 | 65±3 |
| SiC-A2-1-11 | 380 | 1620±90 | 34±1 |
| SiC-A2-1-29 | 260 | 11,340±510 | 7.2±0.3 |
| SiC-A2-1-29b | 540 | 300±040 | 69±2 |
| SiC-A2-2-4 | 340 | 1,595±170 | 34±2 |
| SiC-A2-2-9 | 315 | 2,720±310 | 24±2 |
| C-A3-1-6 | 185 | 830±100 | 49±3 |
| C-A3-1-31 | 170 | -410±70 | 151±20 |
| SiC-A3-2-18 | 195 | 3,410±260 | 20±1 |

[a] Grain name indicates phase (SiC or carbonaceous) and NanoSIMS imaging area (A1, A2, A3)
[b] Corrected for 120-nm beam broadening

Table 3. O-anomalous presolar grains in NWA 5958.

| Grain[a] | Diameter (nm) | Group | $^{17}O/^{16}O$ ($\times 10^{-4}$, ±1σ) | $^{18}O/^{16}O$ ($\times 10^{-3}$, ±1σ) |
|---|---|---|---|---|
| A1-1-4 | 380 | 1 | 5.83 ± 0.23 | 1.672 ± 0.043 |
| A1-2-15 | 290 | 1 | 6.57 ± 0.32 | 2.004 ± 0.056 |
| A2-1-3 | 275 | 1 | 6.84 ± 0.38 | 1.906 ± 0.065 |
| A2-1-4 | 315 | 1 | 13.78 ± 0.48 | 1.978 ± 0.056 |
| A2-1-5 | 275 | 1 | 7.76 ± 0.44 | 2.055 ± 0.070 |
| A2-1-5b | 400 | 1 | 5.47 ± 0.28 | 1.996 ± 0.052 |
| A2-1-6 | 275 | 1 | 7.41 ± 0.45 | 2.006 ± 0.073 |
| A2-1-10 | 260 | 4 | 4.11 ± 0.32 | 4.085 ± 0.101 |
| A2-1-12 | 290 | 1 | 6.36 ± 0.33 | 1.940 ± 0.059 |
| A2-1-18 | 340 | 4 | 3.37 ± 0.42 | 2.480 ± 0.104 |
| A2-1-29 | 245 | 1 | 11.20 ± 0.53 | 2.055 ± 0.070 |
| A2-2-8 | 260 | 1 | 10.83 ± 0.87 | 1.934 ± 0.116 |
| A2-2-15 | 325 | 1 | 7.38 ± 0.42 | 1.914 ± 0.067 |
| A3-1-8 | 255 | 1 | 5.88 ± 0.43 | 2.105 ± 0.081 |
| A3-1-14a | 215 | 1 | 8.10 ± 0.50 | 1.834 ± 0.077 |
| A3-1-14b | 185 | 1 | 4.84 ± 0.39 | 1.684 ± 0.078 |
| A3-1-15 | 260 | 4[b] | 2.74 ± 0.24 | 1.925 ± 0.053 |
| A3-1-19 | 215 | 1 | 6.39 ± 0.48 | 1.959 ± 0.083 |
| A3-2-14 | 260 | 4[b] | 2.32 ± 0.29 | 1.828 ± 0.065 |
| A3-2-16 | 215 | 4 | 3.32 ± 0.33 | 2.406 ± 0.082 |
| Terrestrial[c] | | | 3.829 | 2.0052 |

[a] Grain name includes NanoSIMS imaging area (A1, A2, A3, Table 1)
[b] Based on the Group re-definition proposed here; these are Group 3 in the classification scheme of Nittler et al. (1994, 1997)
[c] Standard mean ocean water

## Figure Captions

**Figure 1.** Left: Overview (secondary electron micrograph) of analyzed region of NWA 5958 thin section. Right: SEM images of areas A1-A3 analyzed by NanoSIMS (Table 1) with color-composite NanoSIMS images overlain (red: $^{12}$C, green: $^{16}$O, blue: $^{28}$Si). Scale bars on right are 50μm.

**Figure 2.** Example NanoSIMS images of one 20 μm ×20 μm region of Area 2 containing 3 presolar grains. a) $^{16}$O image. b) $^{17}$O/$^{16}$O ratio image (expressed as δ-value: δR=1000·(R/R$_{std}$ - 1), where R$_{std}$ is the normalizing ratio, here 3.829×10$^{-4}$ based on standard mean ocean water). The $^{17}$O-rich presolar grain A2-1-29 is indicated. c) $^{28}$Si and d) $^{13}$C/$^{12}$C ratio map (in logarithmic scale, relative to the solar ratio of 0.011). Two presolar SiC grains (SiC-A2-1-29 and SiC-A2-1-29b) are indicated on panels c) and d) with their enrichment factors above solar indicated.

**Figure 3**. Abundances of presolar grains in individual analysis areas and for the total analyzed area of NWA 5958 matrix (Table 1). One- and two-σ error limits are indicated in the scale bars for the individual points, dark (light)-grey band indicates 1(2)-σ confidence limits for total data sets. a) Presolar O-rich grains. b) Presolar SiC grains.

**Figure 4.** C-anomalous rim on silicate grain in NWA 5958. a) Composite NanoSIMS RGB image with red indicating $^{12}$C, green indicating $^{16}$O and blue indicating $^{28}$Si distributions. b) False-color NanoSIMS map of δ$^{13}$C values. c) Secondary electron image of same area; cross indicates location of EDS spectrum shown in d), acquired with a 5 keV electron beam.

**Figure 5.** a) O 3-isotope plot of presolar grains (filled circles) identified in NWA 5958 compared to literature data for presolar oxides (diamonds) and silicates (squares). Shaded regions indicate grain Groups defined for presolar oxides (Nittler et al. 1997), modified as discussed in the main text. b) A subset of data from a) plotted on linear scales. Stars indicate compositions of regions of a 15M$_\odot$ Type II supernova (Woosley and Heger 2007), with long-dashed lines indicating mixing of these zones with the He/C zone (off-scale at $^{18}$O/$^{16}$O=1.6, $^{17}$O/$^{16}$O=1.4×10$^{-4}$). The dot-dashed lines indicate mixing of $^{16}$O-rich zones with two mixtures of the H envelope and He/C zone; "Mix-

95" means a 95:5 envelope:He/C mixture). The dotted lines indicate terrestrial ratios in both panels. Literature data are taken from a large number of sources (see, e.g., Nittler et al. 2008; Zinner 2014; Floss and Haenecour 2016a).

**Figure 6**. Bright-field-STEM image (left) and EDS net count elemental maps (right) of NWA 5985 matrix. The matrix is similar to that of CM2 meteorites, i.e., dominated by serpentines in both fern and platy morphologies, with some olivine (ol), and Fe,Ni sulfides with variable Fe:Ni and levels of oxidation, as isolated grains and in Type I PCP assemblages. Chromite (Chr) and Fe, Ni phosphide (Phos) are also present. Bright white areas are voids (pores).

**Figure 7.** a) Bright-field TEM image of presolar grain A2-1-5. The grain shows alternating stripes typically associated with ortho/clinoenstatite layers. b) Bright-field STEM image of grain A2-1-5 and surrounding matrix with abundant phyllosilicates. The white square outline indicates the location of the grain and approximate field of view of a). The scale bar is 1 μm. c) FFT diffractrogram of a high-resolution TEM image of grain A2-1-5. d) STEM-EDS spectrum of grain A2-1-5. The Mo peak is from the support grid.

**Figure 8.** (Top) Bright-field STEM image of presolar grain A2-2-15 (pink arrow) and surrounding matrix. The dashed red box indicates the area of the STEM-EDS mapping. (Middle) EDS spectrum of A2-2-15 extracted from the spectrum image maps, with a composite RGB overlay of AlSiMg on a STEM HAADF image. The Si and Ca counts are attributed to surrounding matrix materials rather than the presolar grain; Cs is from the SIMS measurements. (Bottom) STEM-EDS net count K edge elemental maps.

**Figure 9**. Abundances of presolar silicates and oxides plotted against those of presolar SiC for NWA 5958 and other primitive meteorites; error bars are 1σ. A77=Allan Hills 77307. LAP=La Paz Icefield. GRA=Grave Nunataks. GRV=Grove Mountain. Data sources: (Floss and Stadermann 2009a, 2012; Vollmer et al. 2009; Nguyen et al. 2010; Leitner et al. 2012, 2019; Zhao et al. 2013; Davidson et al. 2014; Haenecour et al. 2018; Nittler et al. 2018, Verdier-Paoletti et al. 2019). All silicate and oxide abundances are for meteorite matrix except for the CM FGRs data point, which is the average of abundances in fine-grained chondrule rims from six meteorites (Leitner et al. 2019).

Figure 1

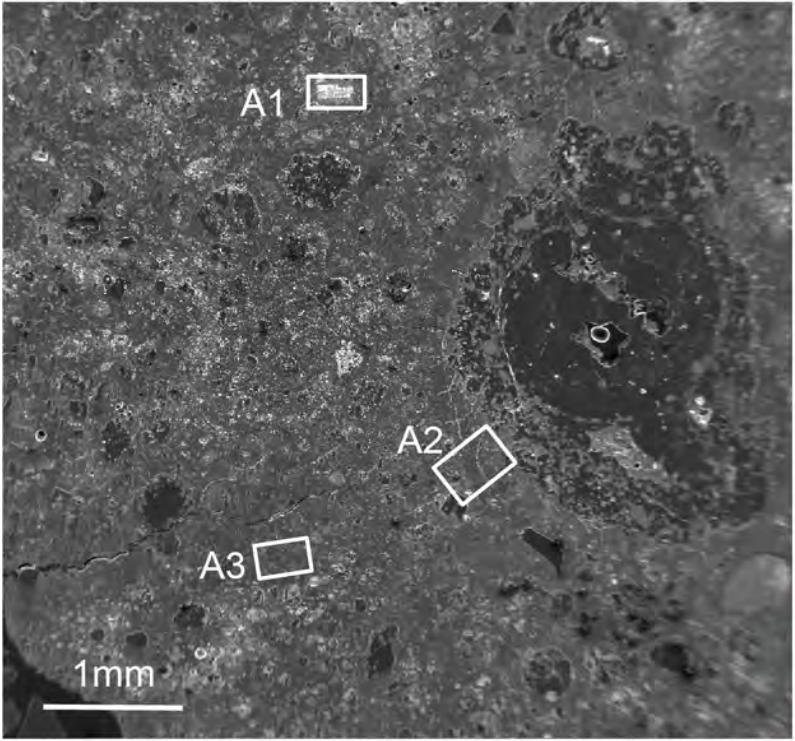
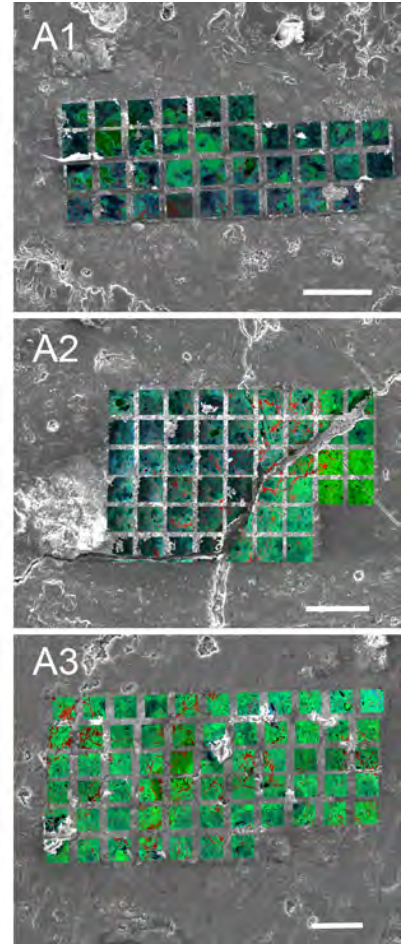

Figure 2

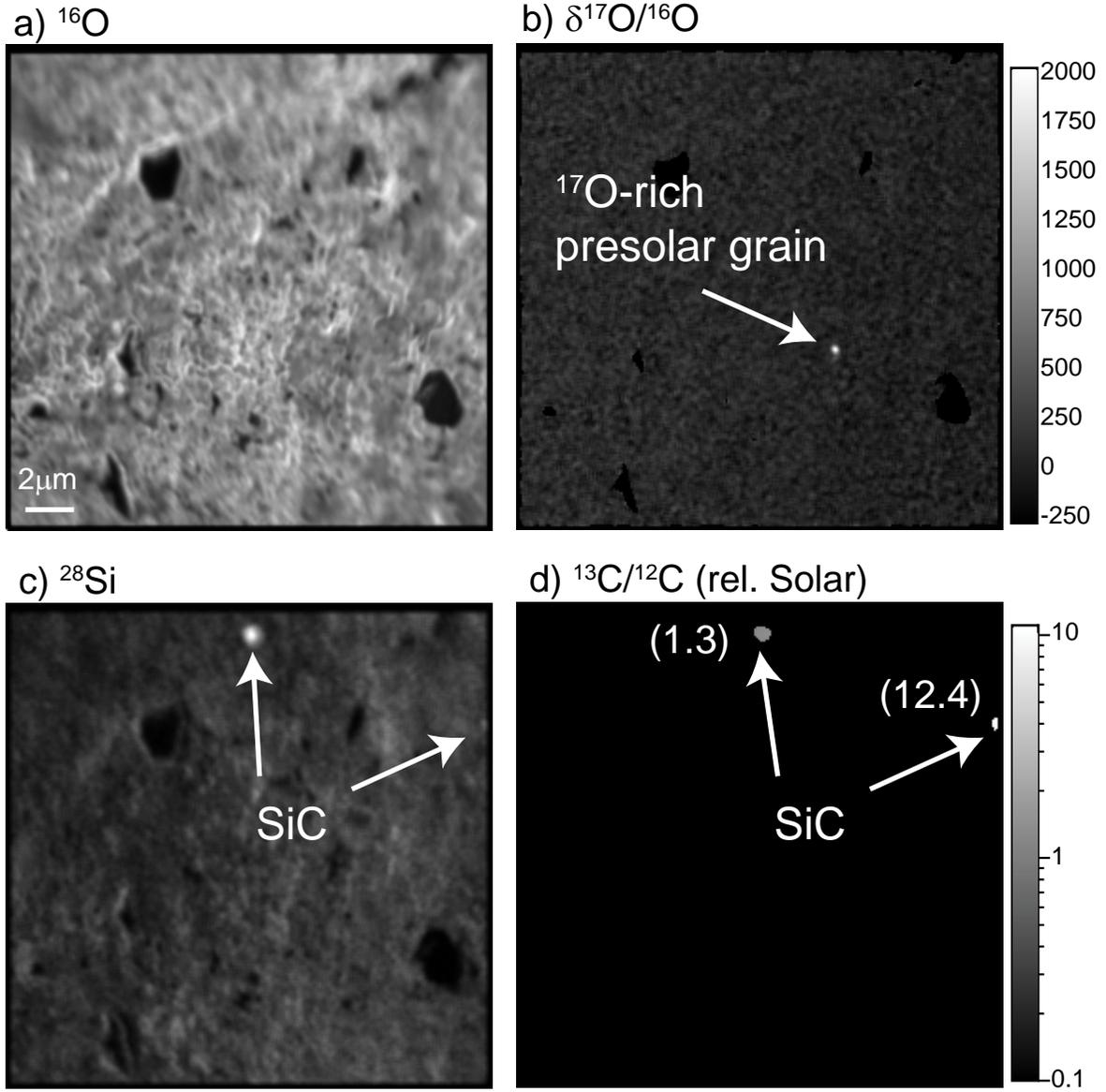

Figure 3

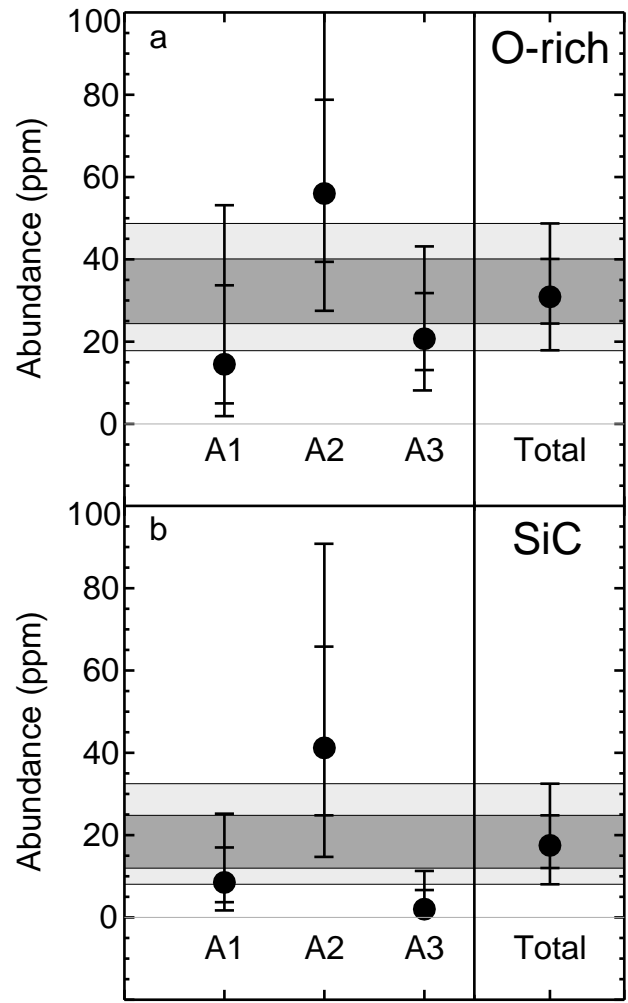

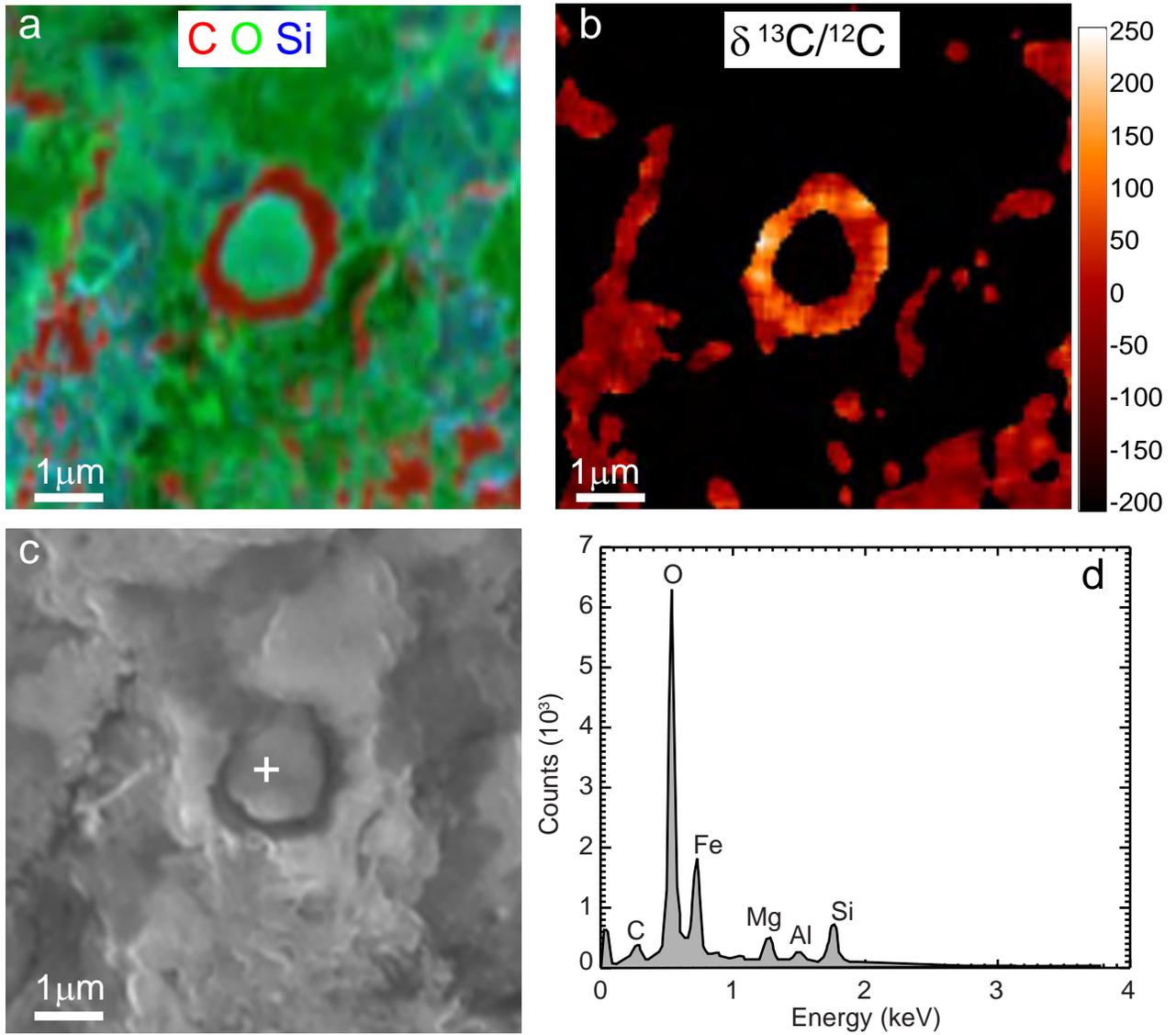

Figure 5

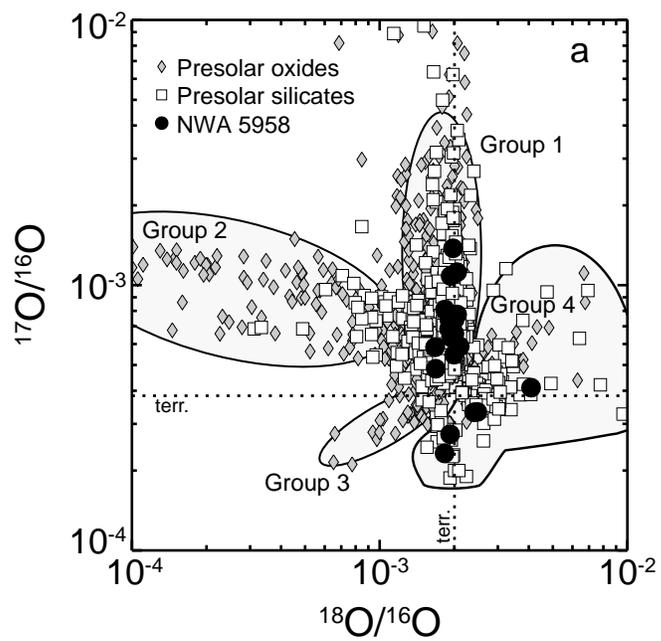

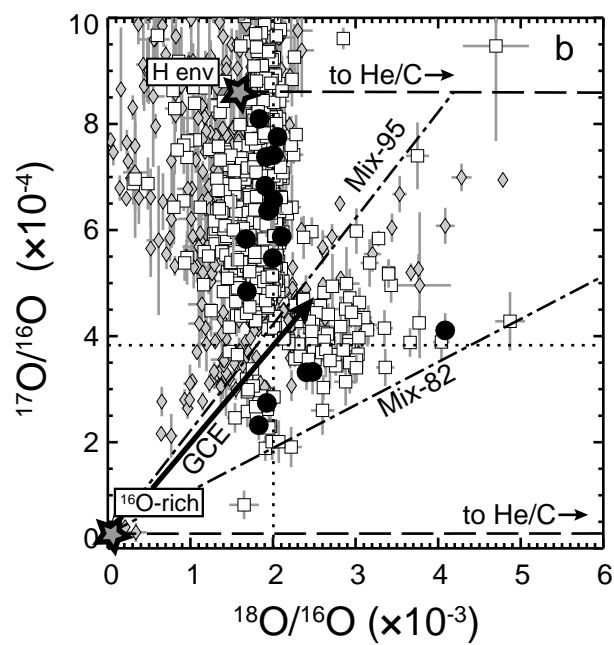

Figure 6

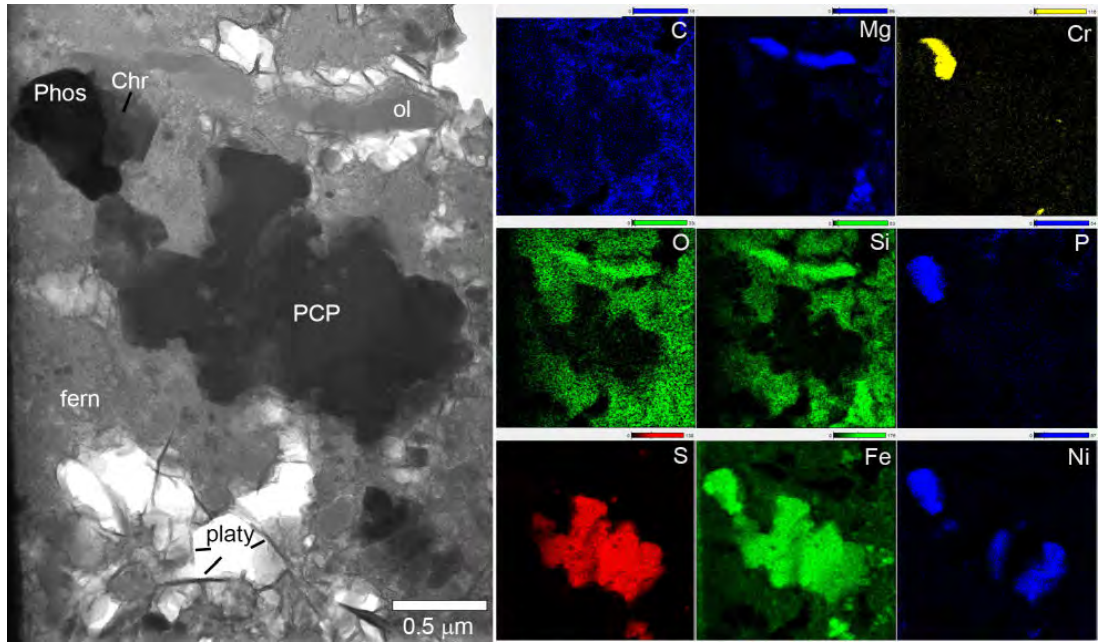

Figure 7

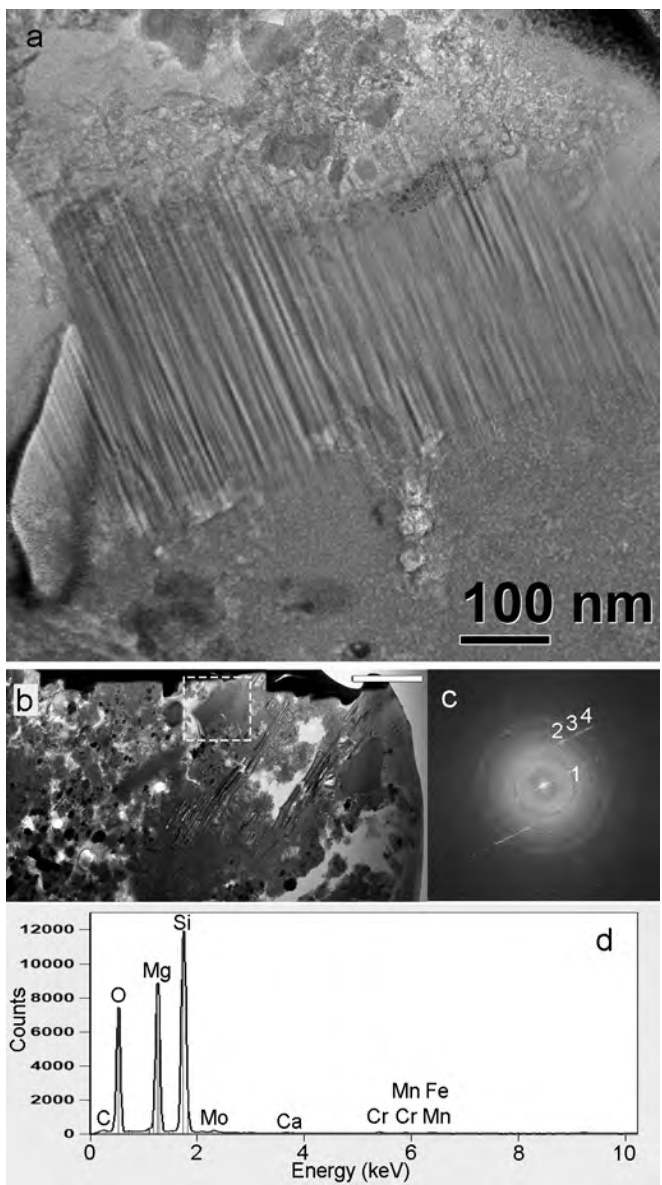

Figure 8

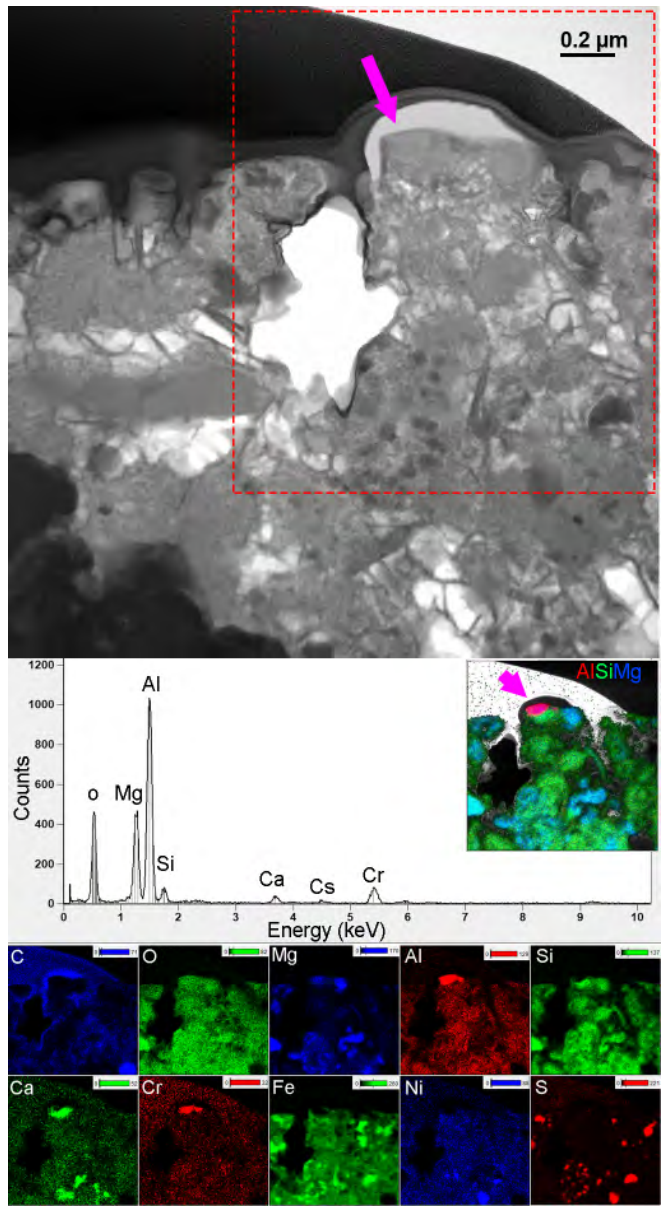

Figure 9

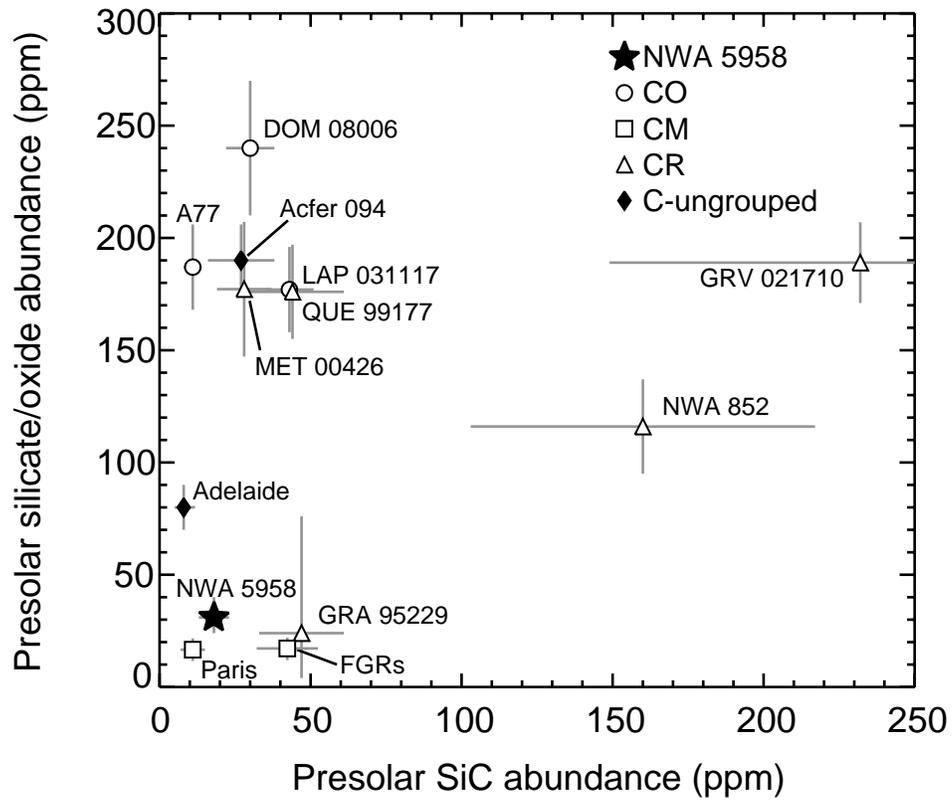